\documentclass[sigconf,anonymous=false]{acmart}
\usepackage{graphicx} 
\usepackage{subcaption}
\usepackage{amsmath} 

\usepackage{amssymb}
\usepackage[linesnumbered,ruled,vlined]{algorithm2e}
\usepackage{array}
\usepackage{xcolor}
\usepackage{booktabs}
\usepackage{pgfplots}
\usepackage[noend]{algpseudocode}
\usepackage{csvsimple}

\pgfplotsset{compat=1.17}

\AtBeginDocument{%
  }

\setcopyright{acmlicensed}
\copyrightyear{2018}
\acmYear{2018}
\acmDOI{XXXXXXX.XXXXXXX}
\acmConference[Conference acronym 'XX]{Make sure to enter the correct
  conference title from your rights confirmation emai}{July 13--18,
  2025}{Woodstock, NY}
\acmISBN{978-1-4503-XXXX-X/18/06}

\copyrightyear{2025}
\acmYear{2025}
\setcopyright{cc}
\setcctype{by}
\acmConference[RecSys '25]{Proceedings of the Nineteenth ACM Conference on Recommender Systems}{September 22--26, 2025}{Prague, Czech Republic}
\acmBooktitle{Proceedings of the Nineteenth ACM Conference on Recommender Systems (RecSys '25), September 22--26, 2025, Prague, Czech Republic}\acmDOI{10.1145/3705328.3748052}
\acmISBN{979-8-4007-1364-4/2025/09}

\begin{document}

\title{RecPS: Privacy Risk Scoring for Recommender Systems }

\author{Jiajie He}
\authornote{Both authors contributed equally to this research.}
\email{jiajih1@umbc.edu}
\orcid{0009-0009-7956-8355}
\affiliation{%
  \institution{University of Maryland, Baltimore County}
  \city{Baltimore}
  \state{Maryland}
  \country{USA}
}

\author{Yuechun Gu}
\authornotemark[1]
\email{ygu2@umbc.edu}
\affiliation{%
  \institution{University of Maryland, Baltimore Country}
  \city{Baltimore}
  \country{USA}
  }

\author{Keke Chen}
\email{kekechen@umbc.edu}
\affiliation{%
  \institution{University of Maryland, Baltimore Country}
  \city{Baltimore}
  \country{USA}
  }

\begin{abstract}
Recommender systems (RecSys) have become an essential component of many web applications. The core of the system is a recommendation model trained on highly sensitive user-item interaction data. While privacy-enhancing techniques are actively studied in the research community, the real-world model development still depends on minimal privacy protection, e.g., via controlled access. Users of such systems should have the right to choose \emph{not} to share highly sensitive interactions. However, there is no method allowing the user to know which interactions are more sensitive than others. Thus, quantifying the privacy risk of RecSys training data is a critical step to enabling privacy-aware RecSys model development and deployment. We propose a membership-inference attack (MIA)- based privacy scoring method, RecPS, to measure privacy risks at both the interaction and user levels. The RecPS interaction-level score definition is motivated and derived from differential privacy, which is then extended to the user-level scoring method. A critical component is the interaction-level MIA method RecLiRA, which gives high-quality membership estimation. We have conducted extensive experiments on well-known benchmark datasets and RecSys models to show the unique features and benefits of RecPS scoring in risk assessment and RecSys model unlearning. Our code is available at \url{https://anonymous.4open.science/r/RsLiRA-4BD3/readme.md}
\end{abstract}



\received{20 February 2007}
\received[revised]{12 March 2009}
\received[accepted]{5 June 2009}
\maketitle

\section{Introduction}
\label{sec:intro}
Recommender systems (RecSys) utilize machine learning algorithms to analyze user-item interactions (i.e., user implicit preferences on items, such as book reviews on Amazon and movie ratings on Netflix) and recommend items to users who might like them \cite{li2023selective}. They have been deployed in numerous applications, including e-commerce, social media, and entertainment. The success of recommendation systems relies on large-scale user personal data, which often contains private information about user preferences, actions, and social contexts \cite{ge2024survey}, thereby raising significant privacy concerns.

Several studies have been made to ensure the privacy of data contributors in RecSys modeling. However, successful deployments of privacy-enhancing techniques have been limited. Earlier efforts in RecSys privacy protection focused on data anonymization techniques \cite{ge2024survey}. Recently, Federated RecSys \cite{wang2022fast} applied the federated learning framework to RecSys modeling, allowing data contributors to keep their data locally and only upload intermediate computational results, thereby avoiding the exposure of raw private data and the application of flawed anonymization techniques. However, federated RecSys does not protect the learned model. Recent studies \cite{yuan2023interaction,zhong2024interaction,zhang2021membership,wang2022debiasing} show that membership inference attacks (MIAs) can reveal private information in training data by only accessing the RecSys model API. Although differentially private machine learning is a provable method for protecting privacy leakage in RecSys models \cite{mullner2023differential}, it leads to significant reductions in model quality due to noise addition and gradient clipping \cite{mullner2023differential}, which is not well accepted by practitioners.

In practice, recommender system (RecSys) practitioners rely primarily on minimal privacy protections, such as controlled access, which fully preserves data utility. In controlled access settings, data curators and authorized users are trusted to protect data privacy. However, numerous risks such as insider threats, system vulnerabilities, and emerging model-API-based attacks \cite{ge2024survey,deldjoo2021survey,rezaimehr2021survey} demonstrate that controlled access alone is insufficient to ensure robust privacy protection.

Consequently, it is essential that data contributors, who are also users of models, can proactively assess their privacy risks and make informed decisions regarding their participation in RecSys projects. Upon receiving estimated privacy risks, data contributors should have the right to request the removal of their data from training datasets \cite{chen2022recommendation,nguyen2022survey} or require the model to unlearn their records \cite{tarun2023fast,zhang2023review}. Recent privacy regulations, such as GDPR \cite{voigt2017eu} and CCPA \cite{ccpa2018}, emphasize the responsibility of data curators to transparently communicate potential privacy risks and enable contributors to opt out of high-risk activities. Despite this regulatory framework, a lack of formal, quantitative tools remains for systematically and transparently evaluating privacy risks.

\textbf{Scope of Research}. Inspired by recent developments in hypothesis-based membership inference \cite{carlini2022lira}, e.g., the likelihood ratio test (LiRA) method, we design an effective RecSys Privacy Scoring Tool (RecPS) to assess the potential risk of a user participating in a RecSys modeling task. This privacy scoring tool has many potential uses. One critical application is for users to determine whether they want to withdraw samples from a RecSys modeling task or ``unlearn'' \cite{chen2022recommendation} selected samples from an existing RecSys model. 

Our scoring approach consists of two critical components: the theory and the implementation. It is backed by the theory of differential privacy, i.e., how difficult an attacker can distinguish whether a user or their specific record is in the training data of a RecSys model. This is the most fundamental privacy threat -- if the attacker cannot confidently determine the membership, the information collected from any other type of attacks, e.g., data reconstruction and property inference \cite{deldjoo2021survey}, cannot be linked to the specific user. We define the formal scoring method for a single user-item interaction and then extend it to the collection of user-item interactions for a specific user. The core is to estimate the optimal ratio between the True Positive Rate (TPR) and False Positive Rate (FPR) of MIA, TPR/FPR. A previous study \cite{song2021systematic} has used MIA's TPR to represent the privacy score. However, we find that the ratio TPR/FPR directly links to the definition of differential privacy, and thus it's more theoretically justifiable. 

However, there are challenges in applying RecSys MIA to the proposed scoring method. (1) Our scoring method requires a powerful MIA at the interaction level. However, we found that most existing RecSys MIA methods only apply to the user level \cite{zhang2021membership,wang2022debiasing,zhu2023membership} and the interaction-level MIAs \cite{yuan2023interaction,zhong2024interaction} have limited attacking abilities and do not meet the requirement of our scoring framework. (2) A promising record-level MIA method, likelihood-ratio attack (LiRA) \cite{carlini2022lira}, has shown state-of-the-art performance on classification models \cite{carlini21Onion, steinke2024privacy,nasr2023tight}. However, it remains unclear whether and how LiRA can be applied to RecSys models for our scoring purposes.  

We designed the RecLiRA MIA for our scoring framework based on LiRA. RecLiRA shows significantly higher TPR in the low-FPR region for interaction-level MIA compared to the best existing interaction-level MIA, MINER \cite{zhong2024interaction}. It works for models that predict the likelihood of interaction, such as neural collaborative filtering (NCF) \cite{he2017neural}, and light graph convolution network (LightGCN) \cite{he2020lightgcn}.

We have evaluated the proposed method with three well-known benchmark recommendation datasets: Amazon Digital Music (ADM) \cite{mcauley2015image}, the Amazon Beauty \cite{mcauley2015image}, and Movielens-1m (ml-1m) \cite{harper2015movielens}. RecLiRA demonstrates significantly better MIA performance than MINER, thereby ensuring the quality of the privacy scores. We also simulate the scenario where users request to remove their records from the RecSys model and show how user-level removals can significantly impact the model performance. We then demonstrate how interaction-level privacy scores can be utilized to selectively remove sensitive interactions, enabling finer-grained privacy-utility tradeoffs.  

Our contributions can be summarized as follows: 
\begin{itemize}
\item To the best of our knowledge, we are the first to propose and study privacy scoring in RecSys, which can serve as a critical tool for both data contributors and RecSys owners to assess the privacy risks of participating in RecSys modeling. 
\item Our RecPS method can generate privacy scores at both the user level and the user-item interaction level using a high-quality MIA method, RecLiRA, which provides multiple granularities for better privacy-utility tradeoffs.  
\item We have conducted extensive experiments to demonstrate the quality of RecLiRA and the unique features of the RecPS scoring method. 
\end{itemize}

The remaining sections are organized as follows. Section \ref{sec:pre} briefly describes the background knowledge; Section \ref{sec:RecPS} shows the principles and implementations of our designed RecPS in detail; Section \ref{sec:exp} presents our experimental results; Section \ref{sec:related work} summarizes the related work; and finally Section \ref{sec:conclusion} concludes our work.

\section{Preliminaries} \label{sec:pre}
In this section, we introduce the basic definitions and notations in Section \ref{sec:definition} and the basic LiRA method that will be adapted to RecSys for our scoring method in Section \ref{sec:lira}.

\subsection{Definitions and Notations} \label{sec:definition}
A recommender system analyzes the interactions between users, denoted by the user set $U$, and items, denoted by the item set $I$, and recommends items that a user may be interested in. A user-item interaction with $u \in U $ and $i \in I$ is denoted as a 3-tuple $(u, i, r)$, where $r$ is a rating that user $u$ gives to item $i$, or simply $(u, i)$, representing $u$ interacted (e.g., clicked) $i$. For simplicity, we used the interaction definition of $(u, i)$ in this paper. A recommender model takes $(u, i)$ as the input and outputs a recommendation score: $s_{(u, i)} = f((u, i))$. We can recommend $i$ if $s_{u, i}$ is in the top-k recommendation scores for all $i \in I$. The output might also be normalized or interpreted as a probability, $s_{(u, i)} \in [0, 1]$. A threshold $\tau$ is given to convert the output to 1, i.e., $u$ ``hits'' $i$, if $ s_{(u, i)} > \tau$; 0 otherwise.

Existing models employ either the scoring function approach, such as ALS \cite{takacs2012alternating}, or the probability output approach, as seen in NCF \cite{he2017neural} and LGCN \cite{he2020lightgcn}. We have adopted the probability output in our approach to conveniently convert the recommender model to a classification model, i.e., the RecSys will predict if it will recommend $i$ to $u$ or not.

\subsection{Likelihood Ratio Attack (LiRA)}\label{sec:lira}
LiRA is a membership inference attack (MIA), which infers the likelihood of a victim sample being a member of the training data for a target model. It has been applied to classification models, where a labeled dataset $D = \{ (x_i, y_i), i=1..N\}$ is used for training the model, where $x_i$ is a sample and $y_i$ is the task-specific label. The membership inference attack can be formally defined as a hypothesis test about a specific sample $(x, y)$ \cite{carlini2022lira}.
\begin{itemize}
\item $H_0$: the sample (x,y) was not included in the training dataset
\item $H_1$: the sample (x,y) was included in the training dataset.
\end{itemize}
To decide between these two hypotheses, LiRA leverages per-sample likelihood ratio test to determine the membership.
\[
\Lambda(f) = \frac{P(f \mid L_D)}{P(f \mid L_{\neg D})}
\]
where $f$ is the model, $L_D$ represents the distribution of logit transformation of the model's highest confidence score for members, and $L_{\neg D}$ represents the distribution of the highest confidence score for nonmembers. $P(f \mid L_D)$ represents the likelihood of $f$ using the sample in training, after observing the logit transformation of the model's highest confidence score of the target sample. LiRA can be done using online or offline testing methods. The offline method is more cost-effective but of slightly lower quality. We have used the offline method in our approach.

\section{Related work}\label{sec:related work}
\textbf{MIA on RecSys.} The earlier RecSys MIA studies are focused on the user level. Zhang et al. \cite{zhang2021membership} propose the Item-Diff method for inferring membership in a target RecSys by analyzing the similarity between a user's historical interactions and recommended items. The core idea is that, for users in the training set, their historical interactions are likely to be more closely aligned with the items recommended by the system. Wang et al. 
 \cite{wang2022debiasing} propose the DL-MIA framework to improve Item-Diff with a VAE-based encoder and weight estimator to address issues with Item-Diff. Note that these user-level MIA methods cannot be modified to perform interaction-level attacks, and thus cannot be adopted by our scoring method. More recently, Wei et al. \cite{yuan2023interaction} propose an interaction-level membership inference on federated RecSys. However, since users in federated RecSys do not expose their records in the first place, it's impossible to calculate the MIA's TPR and FPR for our scoring purpose. Zhong et al. \cite{zhong2024interaction} propose another interaction-level membership inference on KG-based RecSys, utilizing the similarity matrix between the interacted items and the recommended items. However, the version modified for LGCN and NCF models still performs significantly worse than our RecLiRA. 


\textbf{Unlearning on RecSys models. } Users can request to remove sensitive interactions from the recommender system. A closely related issue is machine unlearning \cite{nguyen2022survey}. Current recommender-model unlearning technologies \cite{chen2022recommendation} operate at the user level, assuming the RecSys owner must completely remove all data associated with the user. Such an approach can significantly degrade the overall system performance and exacerbate cold-start issues. We show that interaction-level privacy risk analysis enables finer-grained removal to ensure both utility and privacy guarantees.

\textbf{Privacy Score. } Training data privacy has been a top concern in AI modeling. While methods like differentiated private learning \cite{mullner2023differential} allow data contributors to quantify acceptable privacy loss, model utility is often significantly damaged. In practice, controlled data access remains a mainstream method for protecting data privacy in many industrial and research environments. In controlled data access, authorized model builders work in a restricted environment to access sensitive data, which can fully preserve data utility with reduced risk of data leak. However, unlike differential privacy, there is no quantitative measure for individual data contributors to tell their privacy risk
before participating in a machine learning task. Gu et al. \cite{gu2024} first proposed personalized privacy scoring in Fine tuning LLM but did not conduct in-depth research. At the same time, it is still unknown how to design personalized privacy score for other downstream tasks such as recommendation systems.

\section{RecPS: Privacy Scoring for RecSys} \label{sec:RecPS}
In this section, we first introduce the threat model for RecPS in Section~\ref{sec:threat model}. Then, we define the privacy score in Section~\ref{sec:dp_definition} and describe the score estimation method and implement it in Section~\ref{sec:estmate privacy}. 

\subsection{Threat Model}
\label{sec:threat model}
A typical RecSys consists of two primary components: the offline model training and the online recommendation service that processes user requests. In the scoring process, the model owner simulates a relatively powerful adversary to conduct MIA attacks, who can apply the RecSys model to determine the probability of a user-item interaction. However, the adversary cannot access the offline model training component.

\textbf{Adversary's Goal}: The adversary aims to determine whether a particular user's interaction records were included in training the target recommendation model. Successfully inferring the user's presence in the training dataset could reveal sensitive details about the user's historical interactions, directly compromising the user's privacy and increasing the risk of legal and business consequences for the data collector. The adversary can also conduct other attacks, e.g., data reconstruction, to infer information about a subset or the entire training data. However, without MIA to determine the membership of a user or a record, the adversary cannot confidently link the inferred information to a specific user.  

\textbf{Adversary's Knowledge}: The adversary knows the list of candidate items and the model output, i.e., the probability associated with each recommended item. However, the adversary does not have access to the offline model training process and thus lacks knowledge about the exact users or user-item interactions involved in training.

Under this threat model, RecSys owners can proactively quantify the privacy risk associated with each data contributor. This transparency enables data contributors to understand their privacy risks clearly and make informed decisions about their participation. In particular, when contributors exercise their ``right to be forgotten,'' we show that the scores can help determine a more fine-grained sample removal strategy that preserves more utility with a comparable privacy guarantee.

\subsection{Defining Privacy Score in RecSys} \label{sec:dp_definition}
Determining the likelihood that a user's records were used in modeling is a critical step in defining the privacy risk. Without this information, the adversary cannot link the result of any other type of attack to the specific user.  We derive the theoretically justifiable privacy score based on the definition of differential privacy and its link with MIA. Let's start with the basic definition of differential privacy, e.g., the widely adopted relaxed $(\epsilon,\delta)$ - differential privacy.

\textbf{Background: Differential Privacy (DP).} An algorithm $M$ satisfies $(\epsilon, \delta)$-differential privacy if, for any two adjacent datasets $D_0$ and $D_1$ differing by exactly one record, and for all measurable outputs $\mathcal{O}$, the following condition is met:

$$
Pr(M(D_0) \in \mathcal{O}) \leq e^\epsilon Pr(M(D_1) \in \mathcal{O}) + \delta,
$$

where $\epsilon$ denotes the privacy budget and $\delta$ represents a small probability, commonly set as $\delta = 1/N$ for a dataset containing $N$ records. The symmetry between $D_0$ and $D_1$ encapsulates the concept of ``indistinguishability'' governed by $\epsilon$.

The hypothesis testing interpretation of DP \cite{kairouz2015composition} links the probabilities $Pr(M(D_0)\in \mathcal{O})$ and $Pr(M(D_1)\in \mathcal{O})$ to MIA's True Positive Rate (TPR) and False Positive Rate (FPR). Thus, MIAs can serve as auditing tools to verify whether a claimed $(\epsilon, \delta)$-differentially private model meets the condition: $\ln\left(\frac{TPR}{FPR}\right) \leq \epsilon$ for \emph{every sample}. If a sample causes the violation of this inequality, there might be some DP implementation errors that lead to the false claim of the privacy bound \cite{jagielski2020auditing,steinke2024privacy,nasr2023tight}.

\textbf{Interaction-level privacy risk scoring.} Interestingly, since modeling is a statistical process, we have observed that even without a DP noise injection mechanism, a \emph{non-differentially private modeling method} $M_0$, for two datasets $D_0$ and $D_1$ differing by exactly one \emph{known record} $r$, there exists a sample-specific bound $\epsilon_r$, 
\begin{equation}
\label{eq:tpr/fpr}
\ln\left(\frac{Pr(M_0(D_0)\in \mathcal{O})}{Pr(M_0(D_1)\in \mathcal{O})}\right) \leq \epsilon_r,
\end{equation}
where $\epsilon_r$ represents the ``true'' but unknown privacy risk associated with record $r$. Ideally, a perfectly powerful MIA could precisely estimate this risk with $\ln\left(\frac{TPR}{FPR}\right)$. Motivated by this intuition, we formally define the interaction-level privacy risk as follows. 

In RecSys modeling, since a single record corresponds to an interaction, e.g., $ (u, i) $, we call the \emph{interaction-level} privacy risk bound the 
\emph{privacy score}: 
\begin{equation}
\label{eq:interaction-score}
\small
\epsilon_{(u,i)} 
= \sup \ln\frac{\Pr\bigl(M_0(D_{0,(u,i)})\in\mathcal{O}\bigr)}
             {\Pr\bigl(M_0(D_{1,(u,i)})\in\mathcal{O}\bigr)}.
\normalsize
\end{equation}
where $D_{0, (u, i)}$ and $D_{1, (u, i)}$ differ by $(u, i)$. Without loss of generalization, let $D_{0, (u, i)} = D_{1, (u, i)} \cup (u, i)$. Since we don't know whether an MIA is ideal, we try to find the most powerful MIAs to approach the ideal $\epsilon_r$. 

\textbf{User-level privacy risk scoring.}  Users may also have concerns about their overall privacy risk rather than specific interactions. Existing user-level attacks rely on empirical approaches rather than principled methodologies, such as LiRA \cite{carlini2022lira}, which complicates theoretical privacy analyses. To address this challenge, we propose to derive user-level privacy risk scoring directly from interaction-level privacy risk assessments. Let's denote a user $u$'s interaction set, $I_u$, and $D_{0, u}$ and $D_{1, u}$ represent datasets differing by $I_u$ so we can get $D_{0, u} = D_{1, u} \cup I_u$. Then, $Pr(M_0(D_{0, u})\in \mathcal{O}) = \prod_{i\in I_u} Pr(M_0(D_{0, (u, i)})\in \mathcal{O})$ and $Pr(M_0(D_{1, u})\in \mathcal{O}) = \prod_{i\in I_u} Pr(M_0(D_{1, (u, i)})\in \mathcal{O})$. With Eq. \ref{eq:interaction-score}, we can derive the user's privacy risk, $\epsilon_u$, which is bounded by $\sum_{i \in I_u} \epsilon_{u, i}$.  

To avoid overestimating the risk of users who have a large number of interactions, we have used the average of the interaction scores to represent the user-level score.   
\begin{equation}
\label{eq:group}
\epsilon_u  = 1/|I_u|\sum_{(u, i) \in I_u} \epsilon_{(u, i)}.
\end{equation}

\subsection{RecLiRA: Estimating Privacy Score with MIA}
\label{sec:estmate privacy}
We have formally defined the RecPS scoring methods for both interaction-level and user-level privacy risk bounds. As discussed, we can use an MIA's sample-level TPR and FPR to estimate the interaction-level score, but the MIA's quality is essential to the score. However, very few RecSys-specific MIAs can be applied to the interaction level so far \cite{yuan2023interaction,zhong2024interaction}, whose performance is not strong enough for our scoring method. Thus, we developed a new interaction-level MIA – RecLiRA, based on LiRA originally developed for classification models. In experiments, we have shown that RecLiRA outperforms the best interaction-level MIA reported so far \cite{zhong2024interaction}.

According to the previous discussion, if an interaction-level RecSys MIA, $\mathcal{A}$, exists, we can use its TPR and FPR for a specific interaction $(u, i)$, i.e., $\ln(\text{TPR}_{\mathcal{A}, (u, i)}/\text{FPR}_{\mathcal{A}, (u, i)})$, to approximate the defined privacy score, i.e.,
\[
\hat{\epsilon}_{(u, i)} =\ln(\text{TPR}_{\mathcal{A}, (u, i)}/\text{FPR}_{\mathcal{A}, (u, i)}),
\]
where $\hat{\epsilon}_{(u, i)} \leq \epsilon_{(u, i)}$. With the quality of MIA improving, $\hat{\epsilon}_{(u, i)} \rightarrow \epsilon_{(u, i)}$. Correspondingly, we have 
\[
\hat{\epsilon}_u  = 1/|I_u|\sum_{(u, i) \in I_u} \hat{\epsilon}_{(u, i)}. 
\]

Our next objective is to design a high-quality \emph{interaction-level} MIA, RecLiRA, for RecSys. 

\begin{algorithm}[t]
\caption{The preparation stage}
\label{alg:preparation}
\KwIn{Training dataset $D$, number of shadow models $m$}
\KwOut{Shadow datasets $\{\mathcal{S}_j, j=1..m \}$, shadow models $\{M_j, j=1..m\}$, and distribution $\mathcal{N}_{\text{out}}$}
$\phi_{\text{out}} \gets \{\}$\;
\For{$j \in \{1, ... , m\}$}{
    $\mathcal{S}_j \gets$ Random sampling from the training dataset with each sample selected with 0.5 probability\;
    $M_j \gets \text{Train}(\mathcal{S}_j)$ \Comment{Train a shadow model}\;
    \For{$(u, i) \in \mathcal{D} \setminus \mathcal{S}_j$}{
        $\phi_{\text{out}} \gets \phi_{\text{out}} \cup \{\phi(M_j((u, i)))\}$\;
    }
}
\Comment{Estimate the OUT distribution's parameters: $\mu$ and $\sigma$ with $k$ OUT samples' $\phi$ values}
$\mu_{\text{out}} \gets \text{mean}(\phi_{\text{out}})$\;
$\sigma_{\text{out}} \gets \text{var}(\phi_{\text{out}})$\;
\Return $\{\mathcal{S}_j\}$, $\{M_j\}$, $\mu_{\text{out}}$, $\sigma_{\text{out}}$\;
\end{algorithm}

\begin{algorithm}[t]
\caption{ScoreQuery($u$)}
\label{alg:privacyscoring}
\KwIn{Training dataset $D$, shadow datasets $\{\mathcal{S}_j\}$, shadow models $\{M_j\}$ for $j = 1 \ldots m$,  
      OUT distribution $\mathcal{N}_{out}$, and interaction set $I_u$ of user $u$}
\KwOut{Scores: $\{\hat{\epsilon}_{(u, i)}\}_{(u,i)\in I_u}$ and $\hat{\epsilon}_u$}

\For(\tcp*[h]{For each interaction}){$(u, i) \in I_u$}{  
    \For{$j \in \{1 \ldots m\}$}{
        $\Lambda_{(u,i), M_j} \gets 1 - \Pr\bigl(Z > \phi(q)\bigr), \quad Z \sim \mathcal{N}_{out}$\;
        $L_j \gets \begin{cases}
           1 & \text{if }(u, i) \in \mathcal{S}_j,\\
           0 & \text{otherwise}
        \end{cases}$ \tcp*{Ground-truth IN/OUT}
    }

    \tcp{Gather candidate thresholds from OUT models}
    $T \gets \{\}$\;
    \For{$j \in \{1 \ldots m\}$}{
        \If{$(u,i) \notin \mathcal{S}_j$}{
            $T \gets T \cup \{\Lambda_{(u,i), M_j}\}$\;
        }
    }

    \tcp{Find maximum TPR/FPR}
    $\hat{\epsilon}_{(u, i)} \gets 0$\;
    \For{$t \in T$}{
        $P \gets \{\}$ \tcp*{Predicted labels with threshold $t$}
        \For{$j \in \{1 \ldots m\}$}{
            $P_j \gets \begin{cases}
                1 & \text{if }\Lambda_{(u,i), M_j} > t,\\
                0 & \text{otherwise}
            \end{cases}$\;
        }

        (TPR, FPR) $\gets$ Compute TPR and FPR using $\{L\}$ and $\{P\}$\;

        \If{\text{FPR} $\neq 0$ \textbf{and} $\hat{\epsilon}_{(u, i)} < \ln\bigl(\frac{\text{TPR}}{\text{FPR}}\bigr)$}{
            $\hat{\epsilon}_{(u, i)} \gets \ln\Bigl(\frac{\text{TPR}}{\text{FPR}}\Bigr)$\;
        }
    }
}

$\hat{\epsilon}_u \gets \frac{1}{|I_u|} \sum_{(u, i) \in I_u} \hat{\epsilon}_{(u, i)}$\;

\Return $\{\hat{\epsilon}_{(u, i)}\}_{(u,i)\in I_u}$ and $\hat{\epsilon}_u$\;
\end{algorithm}
We consider the most popular RecSys models, such as NCF \cite{he2017neural} and LightGCN \cite{he2020lightgcn}, which output the probability of a user interacting with an item, i.e., $p = M((u, i)), p \in [0, 1]$. It can be conveniently converted into a binary classifier, with a confidence vector $(p, 1-p)$ for probabilities of (``interaction'', ``no interaction''), respectively. With the classifier representation, we are ready to apply LiRA\footnote{For some scoring models, such as matrix factorization \cite{5197422} and ALS \cite{takacs2012alternating}, which generate scores for ranking and cannot be converted to a classification problem, RecLiRA does not apply. We will develop effective MIAs for such approaches in our future work.}. We have adopted the offline version of LiRA for our scoring approach. Our experiments show that this adaptation is quite successful for RecSys models. 

The basic idea of the offline LiRA is that the output of an IN-training sample $r_{IN}$, $M(r_{IN})$, is highly distinguishable from the OUT-training sample's output distribution, $M(r_{OUT})$. Carlini et al. \cite{carlini2022lira} have identified that the logit transformation of the classification model's output confidence, denoted $q$, can be a useful feature for this task. Intuitively, the IN-training sample is much easier to predict for the classifier than an OUT-training sample, with which the IN samples’ highest confidence scores will be much higher. It’s shown \cite{carlini2022lira} that the logit transformation, $\phi(q) = \log( q/(1-q))$ for both IN and OUT samples has Gaussian-like distributions, which can be conveniently used by hypothesis testing. We have used the absolute difference between the predictions: ``interaction'' and ``no interaction'' to define $q$ for the recommender models to gain satisfactory performance.

The offline LiRA assumes that OUT samples' outputs for different models have a similar $\phi(q)$ distribution and thus can be shared across different models, which significantly saves the cost of MIA. The MIA task can be conducted with a one-side hypothesis testing to determine whether the tested sample's $\phi(q)$ is in the OUT distribution. 

\textbf{RecLiRA Scoring Framework.} We name the LiRA attack for RecSys models: RecLiRA. The framework consists of two stages: the offline preparation stage and the online scoring stage. 
The offline stage prepares the shadow models and estimates the OUT $\phi(q)$ distribution and the scoring stage calculates the privacy score on demand for interactions and users.

As we treat the selected RecSys models as binary classification models, the output confidence vector for a record $(u, i)$ is denoted as $(p, 1-p) = M_0((u, i))$.Intuitively, if one sample is an IN-training sample, the difference between $p$ and $1-p$ should be large, while an OUT-training sample should have $p$ close to $1-p$. Thus, we compute the difference between these two as:
$$
q = |p - (1-p)| = |2p-1|
$$
where $q \in \left[0,1\right]$. With a logit transformation:$
 \phi(q) = \log\left(\frac{q}{1-q}\right)$, $\phi(q)$'s distribution is approximately Gaussian. 

In the preparation stage, we first train $m$ shadow RecSys models, \\
$\{M_1,..., M_m\}$ with the corresponding $m$ sample datasets $\{\mathcal{S}_1,..., \mathcal{S}_m\}$ that are generated as follows. For each sample set, each sample in the original dataset $D$ is selected with 0.5 probability. Thus, each interaction $(u, i) \in D$ shows up in about $M/2$ sample sets. For a sample set, $\mathcal{S}_j$, we name any  $(u,i) \in \mathcal{S}_j$ an IN record for shadow model $M_j$; otherwise, an OUT record for $M_j$. RecPS then collects the $\phi(q)$ values of a few OUT-training samples for estimating the Gaussian distribution $\mathcal{N}_{out}$'s parameters. Algorithm \ref{alg:preparation} describes the detailed steps in the offline preparation stage.

In the online scoring stage, for each interaction $(u, i)$, we calculate its $\phi(q_j)$ for each shadow model $M_j$, where $q_j = M_j( (u, i) )$,  and compute the probability $\Lambda$ as:
\begin{equation}\label{eq:lambda}
\Lambda_j= 1 - Pr(Z > \phi(q_j)), \quad Z \sim \mathcal{N}_{out}
\end{equation}
A higher $\Lambda_j$ indicates the target sample is more likely to be an IN-training sample. We can set up a threshold $T$ for $\Lambda$ as the cutoff separating IN and OUT samples: if $\Lambda_j > T$, we predict the interaction as an IN-training sample. In practice, a global threshold, e.g., $T=0.5$, has led to good results in classification \cite{carlini2022lira}, but may not be optimal. To maximize the ratio TPR/FPR for scoring, we have probed the interaction-specific threshold $T_{(u,i)}$ for each interaction $(u,i)$ as Algorithm \ref{alg:privacyscoring} shows. 

In practice, the number of shadow models, $m$, should be greater than $500$ to generate a statistically significant estimation for TPRs and FPRs \cite{jagielski2020auditing, nasr2023tight, steinke2024privacy}. The OUT samples are used to estimate the parameters of $\mathcal{N}_{out}$, with a minimum of 30 samples to ensure reliable parameter estimation \cite{smith2006central}. 

\textbf{Cost analysis.}
The dominating cost of the proposed RecLiRA is shadow model training. Let's use the cost of training one shadow model as the basic cost unit, $T$. In contrast, model application incurs a much lower cost, denoted as $t$. The whole process involves training $m$ models, applying $k$ OUT samples to infer the OUT distribution, and for each sample testing each of the $m$ models. Thus, the offline cost is $O(mT + kt)$, and the online per-sample cost is $O(mt)$. In experiments, we have used $m$ around 500 and $k$ around 30. 

\section{Experiments}\label{sec:exp}
We design experiments to answer the following research questions.
\begin{itemize}
\item (RQ1) The quality of MIA determines the effectiveness of our privacy scoring. We want to know how effective RecLiRA is on benchmark datasets and RecSys models. 

\item (RQ2) As mentioned in Section \ref{sec:intro}, a critical task to comply with privacy laws is removing specific training samples and unlearning them from a trained model. RecPS privacy scores can guide the removal/unlearning process to preserve more utility. We want to understand how the scores may enable finer-grained privacy-utility tradeoffs.
 
\item (RQ3) The dynamic aspect of RecPS privacy scores: we want to verify whether removing interactions (or users) may affect remaining interactions' and users' scores.

\end{itemize}

\begin{figure*}[tbp]
\centering
\begin{subfigure}{0.32\textwidth}
\centering
\begin{tikzpicture}
\begin{axis}[
    width=1.1\textwidth,
    height=1\textwidth,
    xmode=log,
    ymode=log,
    xmin=9e-5, xmax=1,
    ymin=1e-2, ymax=1.1,
    xtick={1e-4,1e-3,1e-2,1e-1,1},
    xticklabels={$10^{-4}$,$10^{-3}$,$10^{-2}$,$10^{-1}$,$10^{0}$},
    ytick={1e-2,1e-1,1},
    yticklabels={$10^{-2}$,$10^{-1}$,$10^{0}$},
    xlabel={FPR},
    ylabel={TPR},
    grid=major,
    legend style={
        font=\tiny,
        draw=none,
        fill=none,
        at={(0.05,0)},
        anchor=south west,
        text opacity=1
        },
]

\addplot[
  mark=none, 
  color=red,
  line width=1pt, 
  unbounded coords=jump,
]
  table[x=FPR,y=TPR,col sep=comma] {item_LGCN_ml1m_downsampled.csv};
\addplot[
    only marks, 
    mark=*, 
    mark options={color=red, scale=0.7}, 
    forget plot
]
  coordinates {
    (0.000101445,0.333671914)
    (0.001000523,0.348375023)
    (0.010001253,0.478667635)
    (0.100002586,0.769195748)
  };
\addlegendentry{RecLiRA-LGCN(AUC=0.9430)}
\addplot[
  mark=none, 
  color=blue,
  line width=1pt,
  unbounded coords=jump
]
  table[x=FPR,y=TPR,col sep=comma] {item_NCF_ml1m_downsampled.csv};
\addplot[
    only marks, 
    mark=square, 
    mark options={color=blue, scale=0.7}, 
    forget plot
]
  coordinates {
    (0.000101445,0.277603914)
    (0.001000523,0.288440153)
    (0.010001253,0.416131011)
    (0.100002586,0.708664968)
  };
\addlegendentry{RecLiRA-NCF(AUC=0.9169)}
\addplot[
  mark=none, 
  color=orange,
  line width=1pt,
  unbounded coords=jump
]
  table[x=FPR,y=TPR,col sep=comma] {ML_CIKM_sampled.csv};
\addplot[
    only marks, 
    mark=*, 
    mark options={color=orange, scale=0.7}, 
    forget plot
]
  coordinates {
    (0.000105414,0.037099947)
    (0.001030717,0.162017499)
    (0.01004363,0.354381541)
    (0.100043923,0.625148114)
  };
\addlegendentry{MINER-LGCN(AUC=0.8310)}
\addplot[
  mark=none, 
  color=green,
  line width=1pt, 
  unbounded coords=jump
]
  table[x=FPR,y=TPR,col sep=comma] {MN_CIKM_sampled.csv};
\addplot[
    only marks, 
    mark=square, 
    mark options={color=green, scale=0.7}, 
    forget plot
]
  coordinates {
    (0.000111271,0.041144985)
    (0.001030717,0.147570934)
    (0.010037773,0.391329726)
    (0.100248895,0.700363645)
  };
\addlegendentry{MINER-NCF(AUC=0.8727)}
\end{axis}
\end{tikzpicture}
\caption*{\centering (a)MovieLens-1M}
\label{subfig:interaction-tpr-fpr-ML-1M}
\end{subfigure}
\hfill
\begin{subfigure}{0.32\textwidth}
\centering
\begin{tikzpicture}
\begin{axis}[
    width=1.1\textwidth,
    height=1\textwidth,
    xmode=log,
    ymode=log,
    xmin=9e-5, xmax=1,
    ymin=1e-2, ymax=1.1,
    xtick={1e-4,1e-3,1e-2,1e-1,1},
    xticklabels={$10^{-4}$,$10^{-3}$,$10^{-2}$,$10^{-1}$,$10^{0}$},
    ytick={1e-2,1e-1,1},
    yticklabels={$10^{-2}$,$10^{-1}$,$10^{0}$},
    xlabel={FPR},
    ylabel={TPR},
    grid=major,
    legend style={
        font=\tiny,
        draw=none,
        fill=none,
        at={(0.05,0)},
        anchor=south west,
        text opacity=1
        },
]
\addplot[
  mark=none, 
  color=red,
  line width=1pt,
  unbounded coords=jump,  
]
  table[x=FPR,y=TPR,col sep=comma] {item_LGCN_music_downsampled.csv};
\addplot[
    only marks, 
    mark=*, 
    mark options={color=red, scale=0.7}, 
    forget plot
]
  coordinates {
    (0.000104349,0.705541846)
    (0.0010174,0.841984352)
    (0.010017478,0.948229704)
    (0.105627005,1)
  };
\addlegendentry{RecLiRA-LGCN(AUC=0.9988)}
\addplot[
  mark=none, 
  color=blue,
  line width=1pt,
  unbounded coords=jump, 
]
  table[x=FPR,y=TPR,col sep=comma] {item_NCF_music_downsampled.csv};
\addplot[
    only marks, 
    mark=square, 
    mark options={color=blue, scale=0.7}, 
    forget plot
]
  coordinates {
    (0.00010434873346724755,0.6644403249550188)
    (0.0010174001513056634,0.8354233684095742)
    (0.010017478412855763,0.9560820020718609)
    (0.10562700545222133,0.9994002508042091)
};
\addlegendentry{RecLiRA-NCF(AUC=0.9984)}
\addplot[
  mark=none, 
  color=orange,
  line width=1pt, 
  unbounded coords=jump, 
]
  table[x=FPR,y=TPR,col sep=comma] {DL_CIKM_sampled.csv};
\addplot[
    only marks, 
    mark=*, 
    mark options={color=orange, scale=0.7}, 
    forget plot
]
  coordinates {
    (0.0001,0.0265904)
    (0.001005725,0.035479632)
    (0.010057249,0.241400634)
    (0.100030945,0.576486048)
};
\addlegendentry{MINER-LGCN(AUC=0.8449)}
\addplot[
  mark=none, 
  color=green,
  line width=1pt,
  unbounded coords=jump
]
  table[x=FPR,y=TPR,col sep=comma] {DN_CIKM_sampled.csv};
\addplot[
    only marks, 
    mark=square, 
    mark options={color=green, scale=0.7}, 
    forget plot
]
  coordinates {
    (0.0001,0.402488985)
    (0.001005725,0.411146324)
    (0.010057249,0.433176161)
    (0.100263036,0.453273556)
};
\addlegendentry{MINER-NCF(AUC=0.4888)}
\end{axis}
\end{tikzpicture}
\caption*{\centering (b)Amazon Digital Music}
\label{subfig:interaction-tpr-fpr-ADM}
\end{subfigure}
\hfill
\begin{subfigure}{0.32\textwidth}
\centering
\begin{tikzpicture}
\begin{axis}[
    width=1.1\textwidth,
    height=1\textwidth,
    xmode=log,
    ymode=log,
    xmin=9e-5, xmax=1,
    ymin=1e-2, ymax=1.1,
    xtick={1e-4,1e-3,1e-2,1e-1,1},
    xticklabels={$10^{-4}$,$10^{-3}$,$10^{-2}$,$10^{-1}$,$10^{0}$},
    ytick={1e-2,1e-1,1},
    yticklabels={$10^{-2}$,$10^{-1}$,$10^{0}$},
    xlabel={FPR},
    ylabel={TPR},
    grid=major,
    legend style={
        font=\tiny,
        draw=none,
        fill=none,
        at={(0.05,0)},
        anchor=south west,
        text opacity=1
        },
]
\addplot[
  mark=none, 
  color=red,
  line width=1pt, 
  unbounded coords=jump, 
]
  table[x=FPR,y=TPR,col sep=comma] {item_LGCN_beauty_downsampled.csv};
\addplot[
    only marks, 
    mark=*, 
    mark options={color=red, scale=0.7}, 
    forget plot
]
  coordinates {
    (0.000104349,0.662822234)
    (0.0010174,0.834023368)
    (0.010017478,0.954582002)
    (0.105627005,0.997845603)
  };
\addlegendentry{RecLiRA-LGCN(AUC=0.9971)}
\addplot[
  mark=none, 
  color=blue,
  line width=1pt,
  unbounded coords=jump, 
]
  table[x=FPR,y=TPR,col sep=comma] {item_NCF_beauty_downsampled.csv};
\addplot[
    only marks, 
    mark=square, 
    mark options={color=blue, scale=0.7}, 
    forget plot
]
  coordinates {
    (0.000104349,0.662822234)
    (0.001166528,0.81901205)
    (0.010017478,0.954582002)
    (0.105627005,0.997845603)
};
\addlegendentry{RecLiRA-NCF(AUC=0.9984)}
\addplot[
  mark=none, 
  color=orange,
  line width=1pt,
  unbounded coords=jump
]
  table[x=FPR,y=TPR,col sep=comma] {BL_CIKM.csv};
\addplot[
    only marks, 
    mark=*, 
    mark options={color=orange, scale=0.7}, 
    forget plot
]
  coordinates {
    (0.0001,0.152468274)
    (0.001037281,0.336552918)
    (0.010006712,0.572530208)
    (0.100067118,0.832594572)
  };
\addlegendentry{MINER-LGCN(AUC=0.9226)}
\addplot[
  mark=none, 
  color=green,
  line width=1pt,
  unbounded coords=jump
]
  table[x=FPR,y=TPR,col sep=comma] {BN_CIKM_sampled.csv};
\addplot[
    only marks, 
    mark=square, 
    mark options={color=green, scale=0.7}, 
    forget plot
]
  coordinates {
    (0.0001,0.043115065)
    (0.001037281,0.053919485)
    (0.010067728,0.0946627)
    (0.100067118189029,0.319570)
  };
\addlegendentry{MINER-NCF(AUC=0.6786)}
\end{axis}
\end{tikzpicture}
\caption*{\centering (c)Amazon Beauty}
\label{subfig:interaction-tpr-fpr-AB}
\end{subfigure}
\caption{Interaction level: TPR vs FPR on RecLiRA vs MINER.}
\label{fig:interaction-tpr-fpr}
\end{figure*}
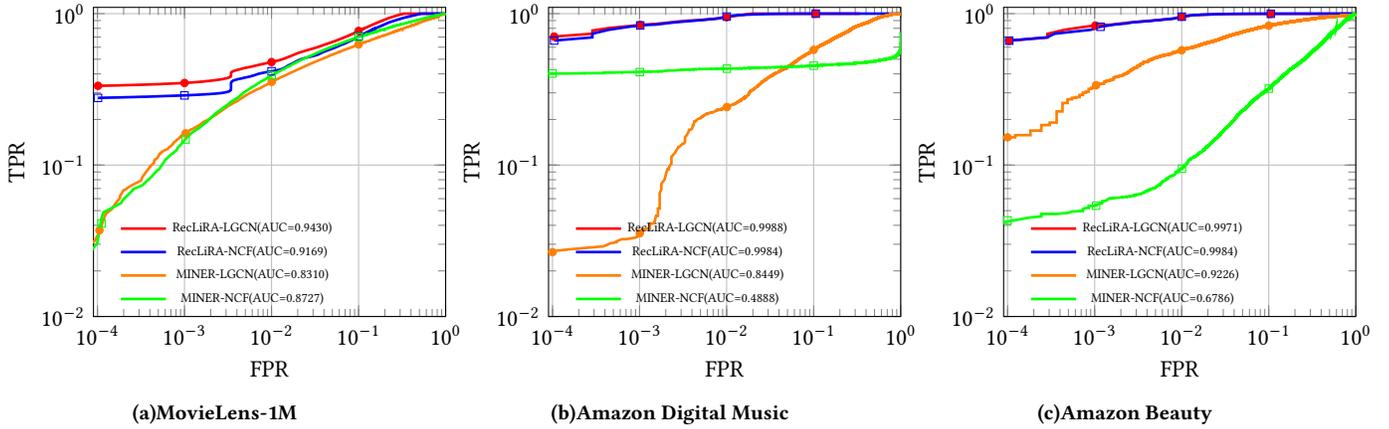

\begin{table}[h]
  \centering
  \resizebox{0.75\linewidth}{!}{%
    \begin{tabular}{l c c c}
      \toprule
      Dataset               & \#Users    & \#Items    & \#Interactions \\
      \midrule
      MovieLens‑1M          & 6,040      & 3,706      & 1,000,209      \\
      Amazon Digital Music  & 840,372    & 456,992    & 1,584,082      \\
      Amazon Beauty         & 1,210,271  & 249,274    & 2,023,070      \\
      \bottomrule
    \end{tabular}%
  }
  \caption{Statistics of datasets.}
  \label{tab:dataset_stats}
\end{table}

\subsection{Experiment Setup}
\textbf{Datasets}. We utilize three real-world datasets in our experiments, including Movielens-1M(ML-1M) \cite{movielens}, Amazon Digital Music (ADM), and Amazon Beauty(AB)  \cite{amazondatasets}, to evaluate our attack strategies. All these datasets are commonly used benchmark datasets for evaluating recommendation systems \footnote{all these datasets are downloaded from \url{https://cseweb.ucsd.edu/~jmcauley/datasets.html} and \url{https://grouplens.org/datasets/movielens/}}. We only keep the users with more than 20 interactions to ensure the performance of the recommender systems \cite{zhang2021membership}. Note that only ratings in these datasets are used for our evaluation in the experiments. Scores range from 1 to 5, which indicates how much users like movies (ML-1M), music (ADM), and beauty or personal care (Amazon Beauty). The statistics of these datasets are shown in Table \ref{tab:dataset_stats}. \ref{tab:dataset_stats}. 

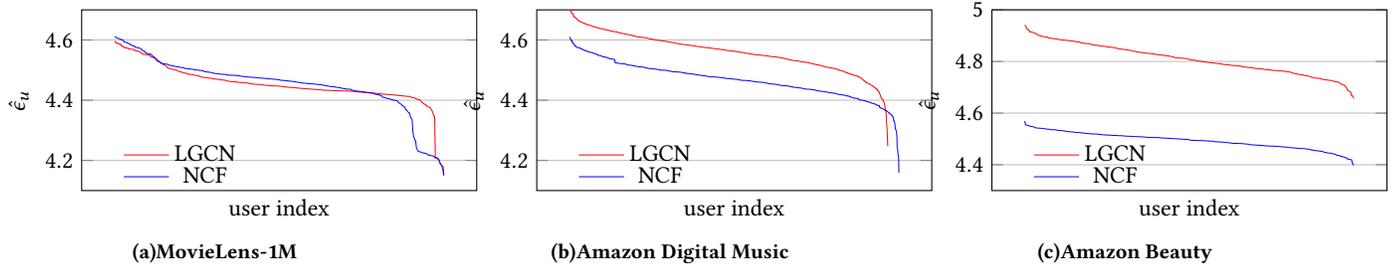
\begin{figure*}[h]
\centering
    \begin{subfigure}[t]{0.32\textwidth}
        \centering
        \begin{tikzpicture}
            \begin{axis}[
                width=1.2\textwidth,
                height=0.7\textwidth,
                xlabel={user index},
                ylabel={$\hat{\epsilon}_u$},
                xtick=\empty,
                ymin=4.1, ymax=4.7,
                grid=major,
                legend style={
                inner sep=-0.2pt,
                outer sep=-0.2pt,     
                fill=none,
                mark size=1pt,
                draw=none,   
                at={(0.4,0)}, 
                anchor=south east,
                row sep=-3pt}   
            ]
            \addplot[color=red, line width=0.3pt] table [
                col sep=comma,
                x expr=\coordindex,
                y={LN(TPR/FPR)-LiRA}
            ] {LGCN_ML1M.csv};
            \addlegendentry{LGCN}
            \addplot[color=blue, line width=0.3pt] table [
                col sep=comma,
                x expr=\coordindex,
                y={LN(TPR/FPR)-LiRA}
            ] {NCF_ML1M.csv};
            \addlegendentry{NCF}
            \end{axis}
        \end{tikzpicture}
        \caption*{\centering (a)MovieLens-1M}
    \end{subfigure}
\hfill
    \begin{subfigure}[t]{0.32\textwidth}
        \centering
        \begin{tikzpicture}
            \begin{axis}[
                width=1.2\textwidth,
                height=0.7\textwidth,
                xlabel={user index},
                ylabel={$\hat{\epsilon}_u$},
                xtick=\empty,
                ymin=4.1, ymax=4.7,
                grid=major,
                legend style={
                inner sep=-0.2pt,
                outer sep=-0.2pt,     
                fill=none,
                mark size=1pt,
                draw=none,   
                at={(0.4,0)}, 
                anchor=south east,
                row sep=-3pt}   
            ]
            \addplot[color=red, line width=0.3pt] table [
                col sep=comma,
                x expr=\coordindex,
                y={LN(TPR/FPR)-LiRA}
            ] {LGCN_music.csv};
            \addlegendentry{LGCN}
            \addplot[color=blue, line width=0.3pt] table [
                col sep=comma,
                x expr=\coordindex,
                y={LN(TPR/FPR)-LiRA}
            ] {NCF_music.csv};
            \addlegendentry{NCF}
            \end{axis}
        \end{tikzpicture}
        \caption*{\centering (b)Amazon Digital Music}
    \end{subfigure}
\hfill
    \begin{subfigure}[t]{0.32\textwidth}
        \centering
        \begin{tikzpicture}
            \begin{axis}[
                width=1.2\textwidth,
                height=0.7\textwidth,
                xlabel={user index},
                ylabel={$\hat{\epsilon}_u$},
                xtick=\empty,
                ymin=4.3, ymax=5,
                grid=major,
                legend style={
                inner sep=-0.2pt,
                outer sep=-0.2pt,     
                fill=none,
                mark size=1pt,
                draw=none,   
                at={(0.4,0)}, 
                anchor=south east,
                row sep=-3pt}   
            ]
            \addplot[color=red, line width=0.3pt] table [
                col sep=comma,
                x expr=\coordindex,
                y={LN(TPR/FPR)-LiRA}
            ] {LGCN_beauty.csv};
            \addlegendentry{LGCN}
            \addplot[color=blue, line width=0.3pt] table [
                col sep=comma,
                x expr=\coordindex,
                y={LN(TPR/FPR)-LiRA}
            ] {NCF_beauty.csv};
            \addlegendentry{NCF}
            \end{axis}
        \end{tikzpicture}
        \caption*{\centering (c)Amazon Beauty}
    \end{subfigure}
\caption{Combined Analysis: user index vs privacy risk score(ln(TPR/FPR))}
\label{fig:user privacy risk}
\end{figure*}

\textbf{Recommender Models.} According to the definition of privacy scores and our implementation of RecLiRA, this method can be applied to any RecSys. In our experiments, we consider two widely used RecSys methods with publicly available implementations for reproducibility: Neural Collaborative Filtering (NCF) \cite{he2017neural} and LightGCN (LGCN) \cite{he2020lightgcn}. For NCF, we utilize the original implementation provided by \cite{he2017neural}. For LightGCN, we configure the model with an embedding dimension of 64 and 3 graph convolution layers. To construct the training and evaluation datasets, we first sort each user's interactions in chronological order according to their timestamps. For every user, we hold out the two most recent interactions: the last interaction is used as the test instance, and the second-to-last interaction is used for validation. All remaining interactions constitute the training set with a negative sampling ratio of 1:4. Model training is performed using stochastic gradient descent (SGD) with a learning rate of 0.001, a batch size of 256, and a maximum of 30 epochs. We apply early stopping if the model's performance does not improve over 5 consecutive epochs. 

The target and shadow models share identical architectures and training strategies, aligning with our threat model. To evaluate MIA performance, we randomly split the entire training dataset into an 8:2 ratio without overlap at the user level, using 80\% of the users to train the target model and 20\% for shadow models. All models are trained on 8 NVIDIA RTX 2080 Ti GPUs.

\textbf{Evaluation measures.} To evaluate the performance of RecSys, we adopt ``the hit ratio at top-$k$ recommended items'' (HR@k) as the metric to evaluate the recommendation performance, where \text{k = 100} is used by previous studies \cite{zhang2021membership, wang2022debiasing}. Specifically, we sort the recommender model's outputs for a set of $(u, i)$ and take the top-k results. These $k$ items intersect with the ground-truth items (i.e., the items the user actually clicked) to calculate HR@k. 

To evaluate the MIA performance, we use AUC (Area under the ROC curve) and TPR at low FPR, which have been commonly used by MIA studies \cite{ho2017detecting, kantchelian2015better, kolter2006learning, carlini2022lira}. 

 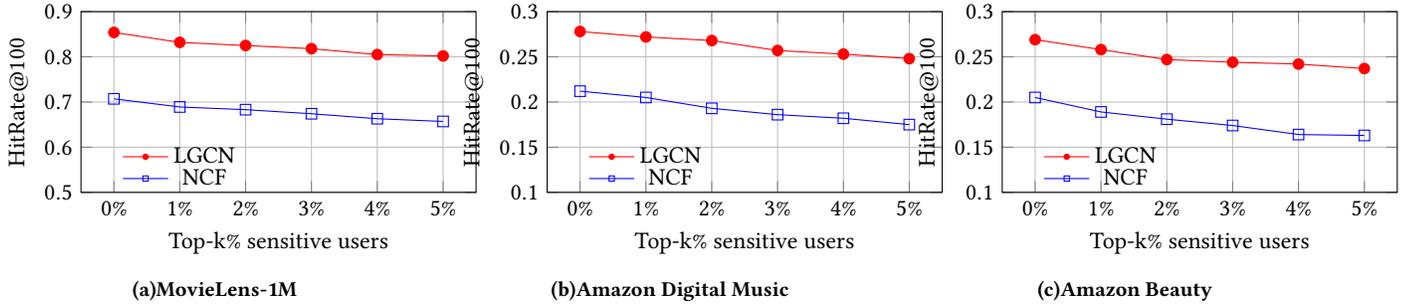
\begin{figure*}[ht]
    \centering
    \begin{subfigure}[t]{0.32\textwidth}
        \centering
        \begin{tikzpicture}
            \begin{axis}[
            	width=1.2\textwidth,
            	height=0.7\textwidth,
                xlabel={Top-k\% sensitive users},
                ylabel={HitRate@100},
                xtick={0,1,2,3,4,5},
                xticklabels={0\%, 1\%, 2\%, 3\%, 4\%, 5\%},
                ymin=0.5, ymax=0.9,
                grid=both,
                legend style={
                inner sep=-0.2pt,
                outer sep=-0.2pt,     
                fill=none,
                mark size=1pt,
                draw=none,   
                at={(0.4,0)}, 
                anchor=south east,
                row sep=-3pt}   
            ]
                \addplot[color=red,mark=*] coordinates {
                    (0, 0.854)
                    (1, 0.832)
                    (2, 0.825)
                    (3, 0.818)
                    (4, 0.805)
                    (5, 0.802)
                };
                \addlegendentry{LGCN}              
                \addplot[color=blue,mark=square] coordinates {
                    (0, 0.707)
                    (1, 0.689)
                    (2, 0.683)
                    (3, 0.674)
                    (4, 0.663)
                    (5, 0.657)
                };
                \addlegendentry{NCF}
            \end{axis}
        \end{tikzpicture}
        \caption*{\centering (a)MovieLens-1M}
        \label{fig:ml1m_users}
    \end{subfigure}
    \hfill
    \begin{subfigure}[t]{0.32\textwidth}
        \centering
        \begin{tikzpicture}
            \begin{axis}[
            	width=1.2\textwidth,
            	height=0.7\textwidth,
                xlabel={Top-k\% sensitive users},
                ylabel={HitRate@100},
                xtick={0,1,2,3,4,5},
                xticklabels={0\%, 1\%, 2\%, 3\%, 4\%, 5\%},
                ymin=0.1, ymax=0.3,
                grid=both,
                legend style={
                inner sep=-0.2pt,
                outer sep=-0.2pt,     
                fill=none,
                mark size=1pt,
                draw=none,   
                at={(0.4,0)}, 
                anchor=south east,
                row sep=-3pt}   
            ]
                \addplot[color=red,mark=*] coordinates {
                    (0, 0.278)
                    (1, 0.272)
                    (2, 0.268)
                    (3, 0.257)
                    (4, 0.253)
                    (5, 0.248)
                };
                \addlegendentry{LGCN}                
                \addplot[color=blue,mark=square] coordinates {
                   (0, 0.212)
                   (1, 0.205)
                   (2, 0.193)
                   (3, 0.186)
                   (4, 0.182)
                   (5, 0.175)
                };
                \addlegendentry{NCF}
            \end{axis}
        \end{tikzpicture}
        \caption*{\centering (b)Amazon Digital Music}
        \label{fig:adm_user}
    \end{subfigure}
    \hfill
    \begin{subfigure}[t]{0.32\textwidth}
        \centering
        \begin{tikzpicture}
            \begin{axis}[
            	width=1.2\textwidth,
            	height=0.7\textwidth,
                xlabel={Top-k\% sensitive users},
                ylabel={HitRate@100},
                xtick={0,1,2,3,4,5},
                xticklabels={0\%, 1\%, 2\%, 3\%, 4\%, 5\%},
                ymin=0.1, ymax=0.3,
                grid=both,
                legend style={
                inner sep=-0.2pt,
                outer sep=-0.2pt,     
                fill=none,
                mark size=1pt,
                draw=none,   
                at={(0.4,0)}, 
                anchor=south east,
                row sep=-3pt}   
            ]
                \addplot[color=red, mark=*] coordinates {
                    (0, 0.269)
                    (1, 0.258)
                    (2, 0.247)
                    (3, 0.244)
                    (4, 0.242)
                    (5, 0.237)
                };
                \addlegendentry{LGCN}                
                \addplot[color=blue,mark=square] coordinates {
                    (0, 0.205)
                    (1, 0.189)
                    (2, 0.181)
                    (3, 0.174)
                    (4, 0.164)
                    (5, 0.163)
                };
                \addlegendentry{NCF}
            \end{axis}
        \end{tikzpicture}
        \caption*{\centering (c)Amazon Beauty}
        \label{fig:ab_users}
    \end{subfigure}
    \caption{Model utility is significantly reduced if sensitive users' interactions are all removed. }
    \label{fig:user_combined_hitrate_analysis}
\end{figure*}

\subsection{Results Analysis}
\textbf{RecLiRA Performance.} We compare RecLiRA with the best performing interaction-level RecSys MIA, MINER\cite{zhong2024interaction}. Considering that MINER was originally designed for knowledge graph (KG)-based recommender systems with a focus on long-tailed distribution and is not directly applicable to NCF and LGCN, we modified certain components of MINER to ensure compatibility with these models. We carefully verified the performance of the modified MINER, specifically focusing on true positive rate (TPR) under low false positive rate (FPR) conditions. Notably, the original MINER paper reported TPR about 0.15 at an FPR of 5\%, while the modified version achieves comparable or superior performance at low FPR levels. Figure \ref{fig:interaction-tpr-fpr} compares RecLiRA and MINER on the interaction-level MIA. The results show that RecLiRA outperforms on all datasets and models. Its overall AUC values are above 0.9 for all the dataset/model combinations. For Music and Beauty datasets, AUC values are almost perfect, around 0.99. More importantly, we have seen high TPRs at the low FPR range ($<0.1$), where high TPR/FPR ratios are observed. In contrast, MINER has a significantly worse AUC and yields lower TPRs in the low FPR range. 

\textbf{User-Level Privacy Score.} To observe how the RecPS scores look like, we also generate user-level scores for 60\% randomly selected users. Figure \ref{fig:user privacy risk} shows the sorted user-level $\ln(\text{TPR}/\text{FPR})$ values. They are all located on the range $[4.06, 4.97]$. The values are approximately divided into three bands. The top 10-20\% has much higher scores, the bottom 10-20\% has much lower, and the middle band shares similar values. Different RecSys models also give different score ranges. LGCN gives higher scores than NCF, seemingly consistent with its better model performance -- LCGN's HR@100 is 10-20\% better than NCF on these datasets.

\begin{figure*}[ht]
    \centering
    
    \begin{subfigure}[t]{0.32\textwidth}
        \centering
        \begin{tikzpicture}
            \begin{axis}[
                width=\linewidth,                      
                height=0.7\textwidth,
                xlabel={Top-k\% sensitive interactions},
                ylabel={HitRate@100},
                xtick={0,10,20,30,40,50, 60, 70},
                xticklabels={0\%, 10\%, 20\%, 30\%, 40\%, 50\%, 60\%, 70\%},
                ymin=0.64, ymax=0.86,
                grid=both,
                legend style={font=\scriptsize, legend columns = -1, fill=none, draw=none, 
                at={(-0.1,1.2)}, anchor=south west},
                x tick label style={font=\tiny}
            ]
                \addplot[color=red, dashed, thick] coordinates {
                    (0, 0.802) (70, 0.802)
                };
                \addlegendentry{remove top-5\% users-LGCN}
                
                \addplot[color=blue, dashed, thick] coordinates {
                    (0, 0.657) (70, 0.657)
                };
                \addlegendentry{remove top-5\% users-NCF}
                
                \addplot[color=red,mark=square,mark size=1.5pt] coordinates {
                    (0, 0.854) (10, 0.8421) (20, 0.839) (30, 0.832) (40, 0.829) (50, 0.823) (60, 0.822) (70, 0.823)
                };
                \addlegendentry{LGCN}
                \addplot[color=red,dashed,mark=*,mark size=1.5pt] coordinates {
                    (0, 0.854) (10, 0.8516) (20, 0.845) (30, 0.84) (40, 0.8374) (50, 0.8341) (60, 0.8304) (70, 0.8277)
                };
                \addlegendentry{LGCN-random}
                \addplot[color=blue,mark=square,mark size=1.5pt] coordinates {
                    (0, 0.707) (10, 0.705) (20, 0.693) (30, 0.687) (40, 0.683) (50, 0.683)(60, 0.673) (70, 0.665)
                };
                \addlegendentry{NCF}
                  \addplot[color=blue,dashed,mark=*,mark size=1.5pt] coordinates {
                    (0, 0.707) (10, 0.695) (20, 0.695) (30, 0.689) (40, 0.686) (50, 0.685)(60, 0.681) (70, 0.677)
                };
                \addlegendentry{NCF-random}
            \end{axis}
        \end{tikzpicture}
        \caption*{\centering (a)MovieLens-1M}
        \label{fig:ml1m_interactions}
    \end{subfigure}
    \hfill
    \begin{subfigure}[t]{0.32\textwidth}
        \centering
        \begin{tikzpicture}
            \begin{axis}[
                width=\linewidth,                      
                height=0.7\textwidth,
                xlabel={Top-k\% sensitive interactions},
                ylabel={HitRate@100},
                xtick={0,10,20,30,40,50, 60, 70},
                xticklabels={0\%, 10\%, 20\%, 30\%, 40\%, 50\%, 60\%, 70\%},
                ymin=0.16, ymax=0.3,
                grid=both,
                legend style={
                font=\tiny,
                inner sep=-0.2pt,
                outer sep=-0.2pt,     
                fill=none,
                mark size=1pt,
                draw=none,   
                at={(1,0)}, 
                anchor=south east,
                row sep=-3pt},
                x tick label style={font=\tiny}
            ]
                \addplot[color=red, dashed, thick] coordinates {
                    (0, 0.248) (70, 0.248)
                };
                
                \addplot[color=blue, dashed, thick] coordinates {
                    (0, 0.175) (70, 0.175)
                };
                \addplot[color=red,mark=square,mark size=1.5pt] coordinates {
                    (0, 0.278) (10, 0.272) (20, 0.265) (30, 0.259) (40, 0.257) (50, 0.253)(60, 0.253) (70, 0.251)
                };
                \addplot[color=red,dashed,mark=*,mark size=1.5pt] coordinates {
                    (0, 0.278) (10, 0.275) (20, 0.272) (30, 0.266) (40, 0.263) (50, 0.258)(60, 0.255) (70, 0.253)
                };
                \addplot[color=blue,mark=square,mark size=1.5pt] coordinates {
                    (0, 0.212) (10, 0.2101) (20, 0.198) (30, 0.192) (40, 0.184) (50, 0.181)(60, 0.180) (70, 0.179)
                };
                 \addplot[color=blue,dashed,mark=*, mark size=1.5pt] coordinates {
                    (0, 0.212) (10, 0.203) (20, 0.195) (30, 0.1933) (40, 0.19) (50, 0.185)(60, 0.182) (70, 0.178)
                };
            \end{axis}
        \end{tikzpicture}
        \caption*{\centering (b)Amazon Digital Music}
        \label{fig:adm_interactions}
    \end{subfigure}
    \hfill
    \begin{subfigure}[t]{0.32\textwidth}
        \centering
        \begin{tikzpicture}
            \begin{axis}[
                width=\linewidth,                      
                height=0.7\textwidth,
                xlabel={Top-k\% sensitive interactions},
                ylabel={HitRate@100},
                xtick={0,10,20,30,40,50, 60, 70},
                xticklabels={0\%, 10\%, 20\%, 30\%, 40\%, 50\%, 60\%, 70\%},
                ymin=0.16, ymax=0.3,
                grid=both,
                legend style={
                font=\tiny,
                inner sep=-0.2pt,
                outer sep=-0.2pt,     
                fill=none,
                mark size=1pt,
                draw=none,   
                at={(1,0)}, 
                anchor=south east,
                row sep=-3pt},
                x tick label style={font=\tiny}
            ]
                \addplot[color=red, dashed, thick] coordinates {
                    (0, 0.237) (70, 0.237)
                };
                
                \addplot[color=blue, dashed, thick] coordinates {
                    (0, 0.163) (70, 0.163)
                };
                \addplot[color=red,mark=square,mark size=1.5pt] coordinates {
                    (0, 0.269) (10, 0.259) (20, 0.251) (30, 0.249) (40, 0.248) (50, 0.243)(60, 0.241) (70, 0.239)
                };
                 \addplot[color=red,dashed,mark=*,mark size=1.5pt] coordinates {
                    (0, 0.269) (10, 0.264) (20, 0.26) (30, 0.256) (40, 0.249) (50, 0.245)(60, 0.243) (70, 0.241)
                };
                \addplot[color=blue,mark=square,mark size=1.5pt] coordinates {
                    (0, 0.205) (10, 0.195) (20, 0.191) (30, 0.191) (40, 0.187) (50, 0.181)(60, 0.179) (70, 0.177)
                };
                \addplot[color=blue,dashed,mark=*,mark size=1.5pt] coordinates {
                    (0, 0.205) (10, 0.197) (20, 0.194) (30, 0.187) (40, 0.186) (50, 0.184)(60, 0.182) (70, 0.18)
                };
            \end{axis}
        \end{tikzpicture}
        \caption*{\centering (c)Amazon Beauty}
        \label{fig:ab_interactions}
    \end{subfigure}
    \caption{Partially removing sensitive users' interactions will preserve RecSys performance better. x-axis: percentage of the top sensitive interactions removed; dash lines are the worst case where all interactions of the target users are removed (the 5\%-users case in Figure 3). }
    \label{fig:combined_hitrate_analysis}
\end{figure*}
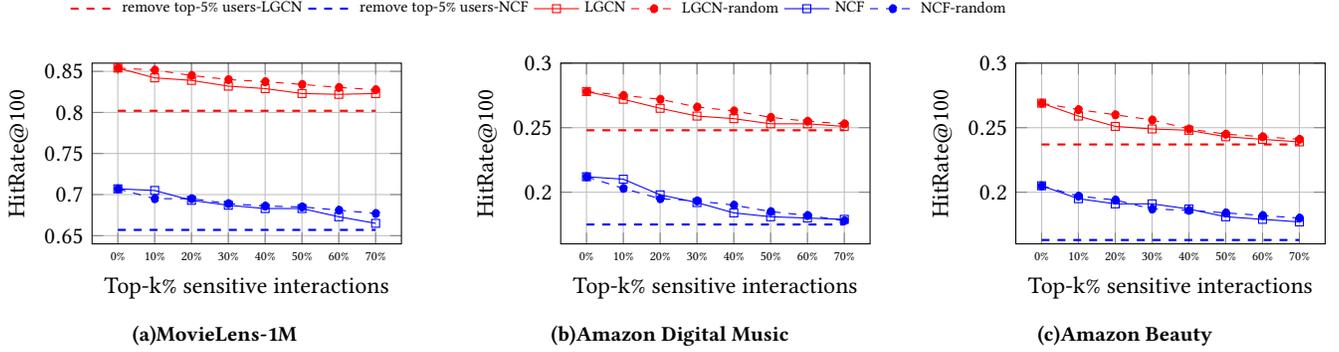

\begin{figure*}[ht]
    \centering
    \begin{subfigure}[t]{0.32\textwidth}
        \centering
        \begin{tikzpicture}
            \begin{axis}[
                width=\linewidth,                      
                height=0.6\textwidth,
                xlabel={Top-k\% sensitive interactions},
                ylabel={Score-reduced Users},
                ylabel style={font=\scriptsize, align=center, xshift=5pt},
                xtick={10,20,30,40,50,60,70},
                xmin = 7 , xmax = 73,
                xticklabels={10\%, 20\%, 30\%, 40\%, 50\%, 60\%, 70\%},
                ymin=0.11, ymax=1.1,
                ytick={0,0.2,0.4,0.6,0.8,1},
                legend style={font=\scriptsize, legend columns = -1, fill=none, draw=none, 
                at={(1,1.2)}, anchor=south west},
                grid=both,
                x tick label style={font=\tiny}
            ]
                \addplot[color=red,mark=square,mark size=1.5pt] coordinates {
                    (10, 0.64)
                    (20, 0.72)
                    (30, 0.89)
                    (40, 0.94)
                    (50, 0.98)
                    (60, 1)
                    (70,1)
                };
                \addlegendentry{LGCN}
                \addplot[color=red,dashed,mark=*,mark size=1.5pt] coordinates {
                    (10, 0.19)
                    (20, 0.32)
                    (30, 0.36)
                    (40, 0.45)
                    (50, 0.53)
                    (60, 0.46)
                    (70,0.45)
                };
                \addlegendentry{LGCN-random}

                \addplot[color=blue,mark=square,mark size=1.5pt] coordinates {
                    (10, 0.54)
                    (20, 0.63)
                    (30, 0.77)
                    (40, 0.93)
                    (50, 1)
                    (60,1)
                    (70,1)
                };
                \addlegendentry{NCF}
               \addplot[color=blue,dashed,mark=*,mark size=1.5pt] coordinates {
                    (10, 0.16)
                    (20, 0.34)
                    (30, 0.45)
                    (40, 0.51)
                    (50, 0.46)
                    (60,0.41)
                    (70,0.53)
                };
                \addlegendentry{NCF-random}
            \end{axis}
        \end{tikzpicture}
        \caption*{\centering (a)MovieLens-1M}
        \label{fig:privacy_preservation_a}
    \end{subfigure}
    \hfill
    \begin{subfigure}[t]{0.32\textwidth}
        \centering
        \begin{tikzpicture}
            \begin{axis}[
                width=\linewidth,                      
                height=0.6\textwidth,
                xlabel={Top-k\% sensitive interactions},
                ylabel={Score-reduced Users},
                ylabel style={font=\scriptsize, align=center, xshift=5pt},
                xtick={10,20,30,40,50,60,70},
                xmin = 7 , xmax = 73,
                xticklabels={10\%, 20\%, 30\%, 40\%, 50\%, 60\%, 70\%},
                ymin=0.11, ymax=1.1,
                legend style={
                    font=\scriptsize,
                    draw=none,
                    fill=none,
                    at={(0,0)},
                    anchor=south west,
                    text opacity=1
                },
                grid=both,
                x tick label style={font=\tiny}
            ]
                \addplot[color=red,mark=square,mark size=1.5pt] coordinates {
                    (10, 0.66)
                    (20, 0.74)
                    (30, 0.87)
                    (40, 0.94)
                    (50, 0.96)
                    (60,0.99)
                    (70,1)
                };
                 \addplot[color=red,dashed,mark=*,mark size=1.5pt] coordinates {
                    (10, 0.21)
                    (20, 0.25)
                    (30, 0.34)
                    (40, 0.45)
                    (50, 0.39)
                    (60,0.41)
                    (70,0.55)
                };

                \addplot[color=blue,mark=square,mark size=1.5pt] coordinates {
                    (10, 0.54)
                    (20, 0.64)
                    (30, 0.69)
                    (40, 0.76)
                    (50, 0.87)
                    (60,0.96)
                    (70,1)
                };
                 \addplot[color=blue,dashed,mark=*,mark size=1.5pt] coordinates {
                    (10, 0.17)
                    (20, 0.21)
                    (30, 0.32)
                    (40, 0.34)
                    (50, 0.35)
                    (60,0.36)
                    (70,0.43)
                };
            \end{axis}
        \end{tikzpicture}
        \caption*{\centering (b)Amazon Digital Music}
        \label{fig:privacy_preservation_b}
    \end{subfigure}
    \hfill
    \begin{subfigure}[t]{0.32\textwidth}
        \centering
        \begin{tikzpicture}
            \begin{axis}[
                width=\linewidth,                      
                height=0.6\textwidth,
                xlabel={Top-k\% sensitive interactions},
                ylabel={Score-reduced Users},
                ylabel style={font=\scriptsize, align=center, xshift=5pt},
                xtick={10,20,30,40,50,60,70},
                xmin = 7 , xmax = 73,
                xticklabels={10\%, 20\%, 30\%, 40\%, 50\%, 60\%, 70\%},
                ymin=0.11, ymax=1.1,
                legend style={
                    font=\scriptsize,
                    draw=none,
                    fill=none,
                    at={(0,0)},
                    anchor=south west,
                    text opacity=1
                },
                grid=both,
                x tick label style={font=\tiny}
            ]
                \addplot[color=red,mark=square,mark size=1.5pt] coordinates {
                    
                    (10, 0.68)
                    (20, 0.75)
                    (30, 0.83)
                    (40, 0.97)
                    (50, 1)
                    (60,1)
                    (70,1)
                };
                \addplot[color=red,dashed,mark=*,mark size=1.5pt] coordinates {
                    (10, 0.25)
                    (20, 0.32)
                    (30, 0.41)
                    (40, 0.45)
                    (50, 0.51)
                    (60,0.56)
                    (70,0.61)
                };

                \addplot[color=blue,mark=square,mark size=1.5pt] coordinates {
                    (10, 0.61)
                    (20, 0.69)
                    (30, 0.75)
                    (40, 0.81)
                    (50, 0.91)
                    (60,0.94)
                    (70,1)

                };
                \addplot[color=blue,dashed,mark=*,mark size= 1.5pt] coordinates {
                    
                    (10, 0.24)
                    (20, 0.26)
                    (30, 0.35)
                    (40, 0.41)
                    (50, 0.37)
                    (60,0.36)
                    (70,0.43)
                };
            \end{axis}
        \end{tikzpicture}
        \caption*{\centering (c)Amazon Beauty}
        \label{fig:privacy_preservation_c}
    \end{subfigure}

    \caption{Removing top-k\% sensitive interactions of the top-5\% sensitive users. Y-axis: the percentage of users whose scores are reduced below the cutoff value $\theta$. The score-guided interaction removals preserve privacy better.}
    \label{fig:interaction-reduction}
\end{figure*}
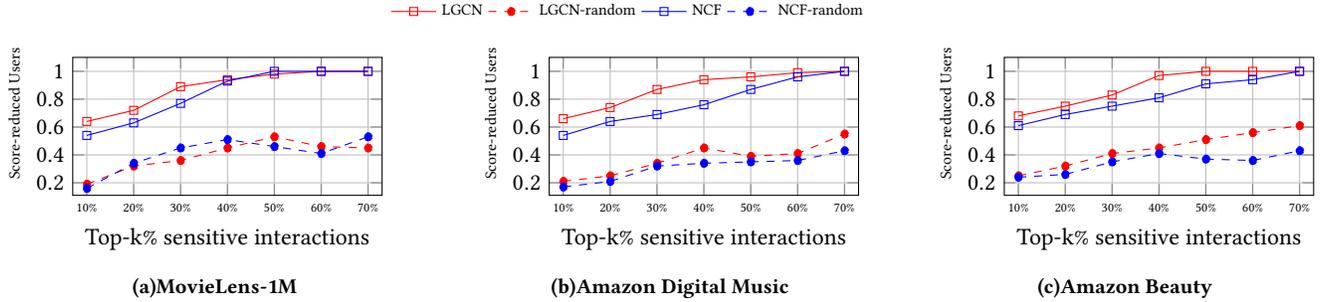

\textbf{How scores affect downstream tasks.}
Users have the right to remove their data from a RecSys model, and according to the laws \cite{voigt2017eu,ccpa2018}, the model owner must ensure that this right is implemented. Consequently, a model retraining or unlearning process \cite{chen2022recommendation} must be applied. We study how model utility is affected in such scenarios and how scores can help both users and model owners make more informed decisions, thereby preserving more model utility. The existing RecSys unlearning strategies \cite{chen2022recommendation} assume that we will remove the entire set of a requested user's interactions, which, however, may lead to a significant decline in the performance of RecSys and cause the cold-start issue for this user. Figure \ref{fig:user_combined_hitrate_analysis} illustrates that with the top-\%1 to top-5\% sensitive users removed based on their privacy scores, HR@100 consistently decreases significantly across all experiments. For instance, on Amazon Digital Music, removing the top 5\% of sensitive users results in a 10.79\% performance drop for LGCN and an even more substantial 37.05\% drop for NCF. 

One may wonder: since some users may have privacy concerns about specific historical interactions but not all, can we remove only the top sensitive interactions to preserve a significantly better model utility, while still protecting privacy satisfactorily? While simulating the actual data removal request stream is impractical in this experiment, we consider an over-simplified case: removing a certain percentage of top-sensitive interactions for the top-sensitive users. Let's define a less aggressive goal. Let the minimum score of the top 5\% users be the cutoff value $\theta$. We consider these users' privacy protection goals to be achieved if their privacy scores (re-evaluated in the new dataset and model) are reduced to below $\theta$ after removing some of their sensitive interactions.  

In Figure \ref{fig:combined_hitrate_analysis}, we show how removing top-sensitive interactions of the top-5\% users affects HR@100. The dashed lines are the model performance if all the top-5\% users' interactions are removed. As expected, we see that better model utility is preserved. 

Figure \ref{fig:interaction-reduction} shows how many of the top-5\% users get their scores reduced below the cutoff $\theta$ with a percentage of their top sensitive interactions removed. We observed less model performance reduction. For example, on Amazon Digital Music dataset, with the top 70\% sensitive interactions removed, 100\% of the top 5\% users successfully get their scores demoted below the cutoff score with only 15.56\% reduction on NCF model performance (in comparison, 37.05\% if removing all interactions); with 70\% of top sensitive interactions removed, 100\% of the top 5\% users successfully get their scores demoted below the cutoff score with only 9.71\% reduction on LGCN model performance (in comparison, 10.79\% if removing all). 

Furthermore, the score-guided interaction removal works much better than random interaction removal. Figure \ref{fig:interaction-reduction} also shows that random removal does not efficiently reduce users' privacy risk. This experiment indicates that the scores can serve as a means for us to fine-tune the data-removal strategies to effectively preserve RecSys utility.

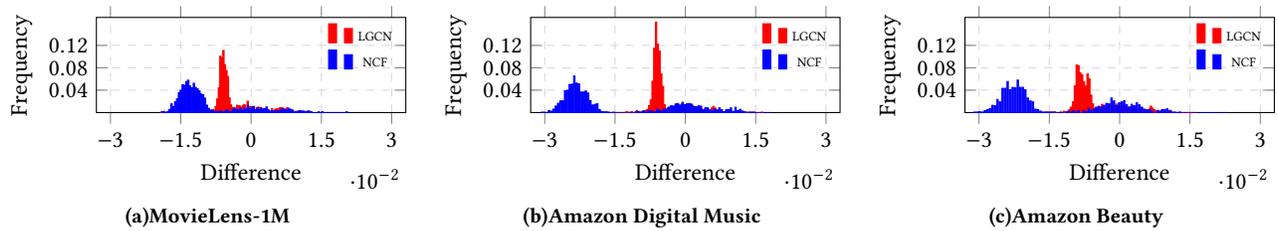
\begin{figure*}[!htb]
    \centering
    \begin{subfigure}[b]{0.32\textwidth}
        \centering
        \begin{tikzpicture}
            \begin{axis}[
                width=\linewidth,                      
                height=0.5\textwidth,                  
                ymin=0, ymax=0.17,                      
                ybar,                                   
                grid=major,                             
                grid style={dashed,gray!30},            
                xmin=-0.03,xmax=0.03,
                xtick={-0.03, -0.015,  0, 0.015, 0.03}, 
                yticklabel style = {/pgf/number format/fixed, /pgf/number format/precision=4},
                ytick={0.04, 0.08,0.12},
                xlabel={Difference},                             
                ylabel={Frequency},                      
                bar width=0.005,                         
                enlarge x limits=0.05,                   
                legend style={
                font=\tiny, 
                fill=none,
                draw=none,   
                at={(1,1)}, 
                anchor=north east}   
            ]
                \addplot+[
                    fill=red,
                    draw=red,
                    bar width = 0.5 pt
                ] 
                table[x=bin_start, y=frequency, col sep=comma] 
                {LGCN_ml1m_user_hist.csv};
                \addlegendentry{LGCN}
                \addplot+[
                    fill=blue,
                    draw=blue,
                    bar width = 0.5 pt
                ] 
                table[x=bin_start, y=frequency, col sep=comma] 
                {NCF_ml1m_user_hist.csv};
                \addlegendentry{NCF}
            \end{axis}
        \end{tikzpicture}
        \label{fig:ncf-ml1m-user}
        \caption*{\centering (a)MovieLens-1M}
    \end{subfigure}
    \begin{subfigure}[b]{0.32\textwidth}
        \centering
        \begin{tikzpicture}
            \begin{axis}[
                width=\linewidth,                      
                height=0.5\textwidth,                  
                ymin=0, ymax=0.17,                      
                xmin=-0.03,xmax=0.03,
                ybar,                                   
                grid=major,                             
                grid style={dashed,gray!30},            
                xtick={-0.03, -0.015,  0, 0.015, 0.03}, 
                yticklabel style = {/pgf/number format/fixed, /pgf/number format/precision=4},
                ytick={0.04, 0.08,0.12},
                xlabel={Difference},           
                ylabel={Frequency},                      
                bar width=0.005,                         
                enlarge x limits=0.05,                   
                legend style={
                font=\tiny, 
                fill=none,
                draw=none,   
                at={(1,1)}, 
                anchor=north east}   
            ]
                \addplot+[
                    fill=red,
                    draw=red,
                    bar width = 0.5 pt
                ] 
                table[x=bin_start, y=frequency, col sep=comma] 
                {LGCN_Music_user_hist.csv};
                \addlegendentry{LGCN}
                \addplot+[
                    fill=blue,
                    draw=blue,
                    bar width = 0.5 pt
                ] 
                table[x=bin_start, y=frequency, col sep=comma] 
                {NCF_Music_user_hist.csv};
                \addlegendentry{NCF}
            \end{axis}
        \end{tikzpicture}
        \label{fig:lgcn-histogram}
        \caption*{\centering (b)Amazon Digital Music}
    \end{subfigure}
    \begin{subfigure}[b]{0.32\textwidth}
        \centering
        \begin{tikzpicture}
            \begin{axis}[
                width=\linewidth,                      
                height=0.5\textwidth,                  
                ymin=0, ymax=0.17,                      
                xmin=-0.03,xmax=0.03,
                ybar,                                   
                grid=major,                             
                grid style={dashed,gray!30},            
                xtick={-0.03, -0.015,  0, 0.015, 0.03}, 
                yticklabel style = {/pgf/number format/fixed, /pgf/number format/precision=4},
                ytick={0.04, 0.08,0.12},
                xlabel={Difference},             
                ylabel={Frequency},                      
                bar width=0.005,                         
                enlarge x limits=0.05,                   
                legend style={
                font=\tiny, 
                fill=none,
                draw=none,   
                at={(1,1)}, 
                anchor=north east}   
            ]
                \addplot+[
                    fill=red,
                    draw=red,
                    bar width = 0.5 pt
                ] 
                table[x=bin_start, y=frequency, col sep=comma] 
                {LGCN_Beauty_user_hist.csv};
                \addlegendentry{LGCN}
                \addplot+[
                    fill=blue,
                    draw=blue,
                    bar width = 0.5 pt
                ] 
                table[x=bin_start, y=frequency, col sep=comma] 
                {NCF_Beauty_user_hist.csv};
                \addlegendentry{NCF}
            \end{axis}
        \end{tikzpicture}
        \label{fig:lgcn-histogram}
        \caption*{\centering (c)Amazon Beauty}
    \end{subfigure}
    \caption{Histagram of users' privacy score difference after removing entire top-5\% sensitive users on each dataset.}
    \label{fig:histogram-user}
\end{figure*}

\textbf{Privacy onion effect.} Since the score-generation process utilizes the training dataset, the change of dataset may affect scores. We have also conducted experiments to investigate the dynamic aspect of RecPS scores. When the top 5\% of sensitive users are removed from the training data, the remaining users' privacy scores change -- the so-called privacy onion effect \cite{carlini21Onion} does exist. Figure \ref{fig:histogram-user} shows the score-change distributions, where x-axis is the score\_difference = new\_score - old\_score. In all experimental settings, we observe the changes are relatively small, within the range $[-0.03, 0.015]$. A small portion of users (around 10-20\%) get scores increased, while most users have scores reduced or unchanged. We observed that interaction-level removal can reduce the privacy onion effect. If we remove only the top 70\% sensitive interactions of the top 5\% sensitive users, the privacy onion effect becomes weaker. The score-increased users are reduced by around 1-7\%. 
The privacy onion effort in the RecPS score evaluation indicates the unique complexity of privacy protection and the interaction-level scoring helps handle this complexity.

\section{Conclusion}\label{sec:conclusion}
Recommender models are built using sensitive user-item interaction data, which may be vulnerable to various model-based attacks. However, current privacy-enhancing techniques are not mature enough to be deployed in recommender systems. Thus, users and model owners must assess the potential privacy risks associated with sharing specific interactions for recommender modeling. We propose the RecPS privacy scoring framework derived from the theory of differential privacy. It can estimate privacy risks at the interaction and user levels with the developed RecLiRA method. Our experimental results demonstrate that the scoring method can generate high-quality scores that effectively guide downstream privacy-enhancing tasks, such as record removal or model unlearning.  
 
\section{Acknowledgment.} This research was partially supported by the National Science Foundation under Grant No. 2517121.

\bibliographystyle{ACM-Reference-Format}
\bibliography{paper}
\end{document}